\documentclass[prd,aps,twocolumn,showpacs, nofootinbib,superscriptaddress,notitlepage]{revtex4-1}
\usepackage{amsmath,graphicx,epsfig,amssymb}
\usepackage{inputenc}
\usepackage[usenames]{color}
\usepackage{slashed}
\usepackage{multirow}
\usepackage{subfigure}
\usepackage{dsfont}
\usepackage{comment}
\usepackage{steinmetz}
\usepackage{setspace}
\usepackage{enumerate}

\allowdisplaybreaks

\usepackage{floatrow}
\newfloatcommand{capbtabbox}{table}[][\FBwidth]

\begin{document}
\title{Revisiting $K_1(1270)- K_1(1400)$ mixing in QCD sum rules}

\author{Yu-Ji Shi}\email{ shiyuji@ecust.edu.cn}
\affiliation{School of Physics, East China University of Science and Technology, Shanghai 200237, China}
\affiliation{INPAC, Key Laboratory for Particle Astrophysics and Cosmology (MOE),  Shanghai Key Laboratory for Particle Physics and Cosmology, School of Physics and Astronomy, Shanghai Jiao Tong University, Shanghai 200240, China}
\author{Jun Zeng}\email{ zengj@sjtu.edu.cn (Corresponding author)}
\affiliation{INPAC, Key Laboratory for Particle Astrophysics and Cosmology (MOE),  Shanghai Key Laboratory for Particle Physics and Cosmology, School of Physics and Astronomy, Shanghai Jiao Tong University, Shanghai 200240, China}
\author{Zhi-Fu Deng}
\affiliation{INPAC, Key Laboratory for Particle Astrophysics and Cosmology (MOE),  Shanghai Key Laboratory for Particle Physics and Cosmology, School of Physics and Astronomy, Shanghai Jiao Tong University, Shanghai 200240, China}

\date{\today}
\begin{abstract}
We investigate the $K_1(1270)-K_1(1400)$ mixing caused by the flavor $SU(3)$ symmetry breaking. The mixing angle is expressed by a $K_{1A}\to K_{1B}$ matrix element induced by the operators that breaks flavor $SU(3)$ symmetry. The QCD contribution to this matrix element is assumed to be dominated and calculated by QCD sum rules. A three-point correlation function is defined and handled both at the hadron and quark-gluon levels. The quark-gluon level calculation is based on operator product expansion up to dimension-5 condensates. A detailed numerical analysis is performed to determine the Borel parameters, and the obtained mixing angle is $\theta_{K_1}=22^{\circ}\pm 7^{\circ}$ or $\theta_{K_1}=68^{\circ}\pm 7^{\circ}$.
\end{abstract}
\maketitle

\section{Introduction}
The flavor $SU(3)$ symmetry plays an important role in the conventional quark model, which classifies hadrons into various of irreducible representations. Although the assumption of perfect flavor $SU(3)$ symmetry succeeds in most of phenomenological analysis on hadron decays and spectrums \cite{Zeppenfeld:1980ex,Savage:1989ub,Deshpande:1994ii,Chau:1986du,Chau:1990ay,Lu:2016ogy,Wang:2017azm,He:2018php}, its breaking effect still cannot be neglected. One of the physical effect due to the flavor $SU(3)$ breaking is the hadron mixing. 

According to the quark model, the two  axial-vector nonets with $J^P=1^+$  are expected as the orbital excitation of the $\bar q q$ system. There are two types of P-wave axial-vector mesons: the $^3 P_1$ and  $^1 P_1$ with the notation $^{2S+1}L_J$. Generally, in the flavor $SU(3)$ limit these two nonets cannot be mixed since they have distinct $C$ parities, explicitly $J^{PC}=1^{++}, \  1^{+-}$ for $^3 P_1,\ ^1 P_1$ respectively. However, because of the mass difference between the strange and light quarks, the kaon nonets $K_{1A}(^3 P_1)$ and $K_{1B}(^1 P_1)$ are distinguished from the mass eigenstates $K_1(1270)$ and $K_1(1400)$. As a result, there emerges a mixing between these two sets of axial-vector kaons: 
\begin{eqnarray}
\begin{pmatrix}
|K_1(1270)\rangle\\
|K_1(1400)\rangle
\end{pmatrix}=
\begin{pmatrix}
\cos\theta_{K_1} & \sin\theta_{K_1}\\
-\sin\theta_{K_1} & \cos\theta_{K_1}
\end{pmatrix}
\begin{pmatrix}
|K_{1B}\rangle\\
|K_{1A}\rangle
\end{pmatrix},
\end{eqnarray}
where $\theta_{K_1}$ is the mixing angle.

Up to now, there are quite a number of studies on $\theta_{K_1}$ in the literature, with most of them based on phenomenological analysis. 
In Ref.\cite{Suzuki:1993yc}, with the use of early experimental information on masses and the partial rates of $K_1(1270)$ and $K_1(1400)$, the authors obtained $\theta_{K_1}=33^{\circ}$ or $57^{\circ}$; The Refs. \cite{Cheng:2003bn,Li:2009tx} phenomenologically analyzed the $\tau \to K_1(1270)\nu_{\tau}$ and $\tau \to K_1(1400)\nu_{\tau}$ decays and obtained $\theta_{K_1}=37^{\circ}$ or $58^{\circ}$; The Ref.\cite{Close:1997nm} studied the correlation of the $f_1(1285)-f_1(1420)$ mixing angle $\theta_{^3 P_1}$ with $\theta_{K_1}$, and obtained $\theta_{K_1}=31.7^{\circ}$ or $56.3^{\circ}$; In Ref.\cite{Cheng:2013cwa}, the authors used the correspondence between $\theta_{K_1}$ and  the  $f_1(1285)-f_1(1420)$, $h_1(1170)-h_1(1380)$ mixing angles to rule out unreasonable $\theta_{K_1}$ values,  and announced a reasonable range as $28^{\circ}<\theta_{K_1}<30^{\circ}$. Besides the phenomenological analysis, in Ref.\cite{Burakovsky:1997dd} the authors obtained $34^{\circ}<\theta_{K_1}<55^{\circ}$ by non-relativistic constituent quark model with the inputs of the mass difference between the $a_1(1260)$ and $b_1(1235)$ mesons, as well as the ratio of the constituent quark masses. The Ref.\cite{Dag:2012zz} related $\theta_{K_1}$ with a two-point correlation function. By calculating this correlation function in QCD sum rules (QCDSR) the authors obtained  $\theta_{K_1}=39^{\circ}\pm 4^{\circ}$.

It can be found that the phenomenological predictions on $\theta_{K_1}$ mentioned above are not unified.  In this work, we desire to give a theoretical and model independent calculation on $\theta_{K_1}$, with the use of QCDSR. It should be mentioned that the QCDSR calculation in Ref.\cite{Dag:2012zz} has a defect. The authors missed the pseudoscalar kaon contribution when operating the $K_{1A}$ current on the correlation function. In this work, we will calculate $\theta_{K_1}$ in a different framework of QCDSR from that in Ref.\cite{Dag:2012zz}. The $\theta_{K_1}$ will be expressed by a $K_{1A}\to K_{1B}$ matrix element induced by $SU(3)$ breaking operators, with zero transferring momentum. This method has been successfully applied to the $\Xi_c-\Xi_c^{\prime}$ mixing \cite{Liu:2023feb,Deng:2023qaf}. The $K_{1A}\to K_{1B}$ matrix element will be calculated in QCDSR by a three-point correlation function. Some introductions on QCDSR and its applications can be found in Refs.~\cite{Onishchenko:2000wf,Onishchenko:2000yp,Kiselev:2001fw,Zhang:2008rt,Wang:2010hs,Wang:2010vn,Wang:2010it,Hu:2017dzi,Hu:2017dzi,Shi:2019hbf}.

This paper is arranged as follows: In section \ref{sec:mixing_angle}, we  introduce the method to extract $\theta_{K_1}$; Section \ref{sec:hadron_level_calculation} gives the hadron level calculation;
Section \ref{sec:QCD_level_calculation} gives the quark-gluon level calculation; Section \ref{sec:Numerical} presents numerical results; Section \ref{sec:summary} is a summary of this work.

\section{$K_1(1270)- K_1(1400)$ mixing}
\label{sec:mixing_angle}
There are two sources of the flavor $SU(3)$ breaking. The first one comes from the mass difference between $s$ and 
$u, d$ (nearly massless) quarks, which only provides QCD contribution to $K_1(1270)- K_1(1400)$ mixing. Another one comes from the electric charge difference among the $u, d, s$ quarks, which involves the QED effect. In this work, we will focus on the QCD contribution since the QED effect is expected to be tiny as shown by our previous work on the $\Xi_c-\Xi_c^{\prime}$ mixing \cite{Liu:2023feb,Deng:2023qaf}.
The full QCD Lagrangian of the quark sector contains both the terms conserving and breaking the flavor $SU(3)$ symmetry: $\mathcal{L}_{\rm QCD}=\mathcal{L}_0+\Delta \mathcal{L}$, where $\mathcal{L}_0$ reads as
\begin{eqnarray}
\mathcal{L}_0&=&\sum_{q}\bar q (i\slashed{D}-m_u) q,
\end{eqnarray}
with $D$ being the QCD covariant derivative, $q=u,d,s$, and $m_u=m_d=0$ is approximately assumed. The $SU(3)$ symmetry breaking term $\Delta \mathcal{L}$, which  arises from the quark mass difference reads as
\begin{eqnarray}
\Delta \mathcal{L}&=&\bar s(m_u-m_s) s.
\end{eqnarray}
Accordingly, the Hamiltonian is decomposed as $H=H_0+\Delta H$, 
with 
\begin{eqnarray}
 \Delta H = \int d^3x \Delta \mathcal{H}(x) = - \int d^3x \Delta \mathcal{L}(x). 
\end{eqnarray}

The lowest axial vector kaons $K_{1}(1270)$ and $K_{1}(1400)$ are the mass eigenstates of the full Hamiltonian $H$:
\begin{align}
&H |K_{1}(1270)\rangle=m_{1270}|K_{1}(1270)\rangle,\nonumber\\
&H |K_{1}(1400)\rangle=m_{1400}|K_{1}(1270)\rangle.\label{eq:physicalMassDef}
\end{align}
On the other hand, in the $SU(3)$ symmetry limit, the lowest axial vector kaons are classified into $K_{1B}(^1 P_1)$ and  $K_{1A}(^3 P_1)$ states, which are eigenstates of the $SU(3)$ conserved Hamiltonian $H_0$:
\begin{align}
&H_0 |K_{1B}\rangle=m_{1B}|K_{1B}\rangle,\nonumber\\
&H_0 |K_{1A}\rangle=m_{1A}|K_{1A}\rangle\rangle.
\end{align}
The mixing between the physical doublet $|K_{P}\rangle=(|K_{1}(1270)\rangle, |K_{1}(1400)\rangle)^T$ and the $SU(3)$ doublet $|K_F\rangle=(|K_{1B}\rangle, |K_{1A}\rangle)^T$ is described by a unitary transforming matrix $U$ with a mixing angle $\theta_{K_1}$:
\begin{eqnarray}
|K_P\rangle=
\begin{pmatrix}
\cos\theta_{K_1} & \sin\theta_{K_1}\\
-\sin\theta_{K_1} & \cos\theta_{K_1}
\end{pmatrix}
|K_F\rangle
=U|K_F\rangle.\label{eq:unitaryTransfer}
\end{eqnarray}

Here we consider the matrix element for the $SU(3)$ doublet $|K_F\rangle$: 
$\langle K_F(p^{\prime})|H|K_F(p)\rangle$. Both the initial and final states are set to be static ${\vec p}={\vec p}^{\prime}=0$ and on-shell $p_{1B}^0=m_{1B},\  p_{1A}^0=m_{1A}$.  With the use of the unitary transformation $U$ defined in Eq.~\eqref{eq:unitaryTransfer}, and the physical masses defined in Eq.~\eqref{eq:physicalMassDef}, we obtain
\begin{align}
&\begin{pmatrix}
\langle K_{1B}(\lambda^{\prime})|H|K_{1B}(\lambda)\rangle &\langle K_{1B}(\lambda^{\prime})|H|K_{1A}(\lambda)\rangle\\
\langle K_{1A}(\lambda^{\prime})|H|K_{1B}(\lambda)\rangle & \langle K_{1A}(\lambda^{\prime})|H|K_{1A}(\lambda)\rangle
\end{pmatrix}\nonumber\\
=&\  2 (2\pi)^3\delta^{(3)}(\vec{0})\delta_{\lambda\lambda^{\prime}}\nonumber\\
\times& \begin{pmatrix}
m_{1270}^2 \ c_k^2 +m_{1400}^2 \ s_k^2 &(m_{1270}^2 -m_{1400}^2) \ c_k s_k\\
(m_{1270}^2 -m_{1400}^2)\ c_k s_k & m_{1270}^2 \ s_k^2+m_{1400}^2 \ c_k^2 
\end{pmatrix},\label{eq:SHSmatrix}\nonumber\\
\end{align}
where $s_k= \sin\theta_{K_1}, c_k= \cos\theta_{K_1}$ for simplicity the momentum dependence  of the $K_{1B, 1A}$ states are not shown in the matrix element.

It can be found that the upper-right off-diagonal component in the Eq.~\eqref{eq:SHSmatrix} leads to the equation 
\begin{eqnarray}
&&\langle K_{1B}(\lambda^{\prime})|H|K_{1A}(\lambda)\rangle\nonumber\\
&=&(2\pi)^3\delta^{(3)}(\vec{0})\delta_{\lambda\lambda^{\prime}}(m_{1270}^2 -m_{1400}^2)\sin 2\theta_{K_1}.\label{eq:1Bto1Amatrix}
\end{eqnarray}
Therefore, the mixing angle $\theta_{K_1}$ can be extracted as soon as one calculate the matrix element on the left hand side above, which can be further expressed as
\begin{eqnarray}
&&\langle K_{1B}(\lambda^{\prime})|H|K_{1A}(\lambda)\rangle\nonumber\\
&=& (2\pi)^3\delta^{(3)}({\vec{0}})\langle K_{1A}(\lambda^{\prime})|\Delta \mathcal{H}(0)|K_{1B}(\lambda)\rangle,\label{eq:1Bto1Amatrix2}
\end{eqnarray}
by integrating out the coordinate. Equating the Eq.~\eqref{eq:1Bto1Amatrix} and Eq.~\eqref{eq:1Bto1Amatrix2}, and  setting $\lambda=\lambda^{\prime}$ we have
\begin{eqnarray}
\sin2\theta_{K_1}=\frac{m_s-m_u}{m_{1270}^2 -m_{1400}^2}\langle K_{1B}|\bar{s}s(0)|K_{1A}\rangle.\label{eq:thetaKformula1}
\end{eqnarray}
Generally, the matrix element $\langle K_{1A}(p_2)|\Delta \mathcal{H}(0)|K_{1B}(p_1)\rangle$ with nonzero initial and final momentums can be parameterized as
\begin{align}
&\langle K_{1B}(p_2)|\bar{s}s(0)|K_{1A}(p_1)\rangle \nonumber\\
=&\  \epsilon^*_\mu(p_2)\bigg[F_1 g^{\mu\nu}+\frac{F_2}{M^2}\epsilon^{\mu\nu\alpha\beta}p_{1\alpha}p_{2\beta}+\frac{F_3}{M^2}p_1^\mu p_2^\nu\bigg]\epsilon_\nu(p_1)\label{eq:1Bto1Apara},
\end{align}
where $F_{1,2,3}$ are three transition functions of $q^2=(p_1-p_2)^2$, and $M^2$ can be the mass of $K_{1A}$ or $K_{1B}$. Note that the matrix element we actually need in Eq.~\eqref{eq:thetaKformula1} requires $p_1=p_2$. Thus the $F_2, F_3$ terms in Eq.~\eqref{eq:1Bto1Apara} vanishes and only $F_1(q^2=0)$ is relevant. Accordingly the mixing angle can be obtained as
\begin{eqnarray}
\sin2\theta_{K_1}=\frac{m_s-m_u}{m_{1400}^2-m_{1270}^2} F_1(0).\label{eq:thetaKformula2}
\end{eqnarray}

\section{Hadron Level calculation}
\label{sec:hadron_level_calculation}
In this and the next section, we will introduce a QCDSR calculation for the matrix element in Eq.~\eqref{eq:1Bto1Apara}. We firstly define a three-point correlation function:
\begin{align}
 \Pi_{\mu\nu\rho}(p_1,p_2)= &\  i^2\int d^4x d^4y\  e^{ip_2\cdot x}e^{-ip_1\cdot y} \nonumber\\
&\times \langle 0|T\{J_{\mu\nu}^{1B}(x)\bar{s}s(0) J_{\rho}^{1A\dagger}(y)\}|0 \rangle,\label{eq:DefineCorrFunc}
\end{align}
where $J_{\mu\nu}^{1A}$ and $J_{\mu\nu}^{1B}$ are the currents of $K_{1A}(^3P_1)$ and $K_{1B}(^1P_1)$, respectively:
\begin{eqnarray}
J_{\mu\nu}^{1B}=\bar{q}\sigma_{\mu\nu}s,~~
J_{\rho}^{1A}=\bar{q}\gamma_\rho\gamma_5 s.\label{eq:kaoncurrent}
\end{eqnarray}
The correlation function defined in Eq.~\eqref{eq:DefineCorrFunc} should be calculated both at the hadron level and the quark-gluon level. At the hadron level, inserting the complete sets with the same quantum number of $K_{1B}$ and $K_{1A}$ into the correlation function, and using the following definition of kaon decay constants:
\begin{eqnarray}
\langle 0|J_{\mu\nu}^{1B}(0)|K_{1B}(p, \lambda) \rangle &=& if_{K_{1B}}^\perp \epsilon_{\mu\nu\alpha\beta}\epsilon^\alpha(p, \lambda) p^\beta, \nonumber\\
\langle 0|J_{\rho}^{1A}(0)|K_{1A}(p, \lambda) \rangle &=& -if_{K_{1A}}m_{K_1A}\epsilon_{\rho}(p, \lambda), \nonumber\\
\langle 0|J_{\rho}^{1A}(0)|K(p) \rangle &=& if_{K}p_\rho,
\end{eqnarray}
one can express the correlation function as
\begin{widetext}
\begin{align}
&\Pi^H_{\mu\nu\rho}(p_1, p_2) \nonumber\\
=& -m_{A}f_{A}f_{B}^{\perp}\epsilon_{\mu\nu\alpha\beta}p_2^\beta \bigg( -g^{\alpha\kappa}+\frac{p_2^\alpha p_2^\kappa}{m_{B}^2} \bigg)\frac{1}{p_2^2-m_{1B}^2}\bigg[  F_1g_{\kappa\tau} + \frac{F_2}{M^2}\epsilon_{\kappa\tau\rho\sigma} p_1^{\rho} p_2^{\sigma} +\frac{F_3}{M^2}p_{1\kappa} p_{2\tau}\bigg]\bigg( -g^{\tau}_\rho+\frac{p_1^\tau p_{1\rho}}{m_{1A}^2} \bigg)\frac{1}{p_1^2-m_{1A}^2}\nonumber\\
&+ f_kf_{1B}^\perp \epsilon_{\mu\nu\alpha\beta} p_2^\beta p_{1\rho}\frac{1}{p_2^2-m_{1B}^2}
\bigg( -g^{\alpha\kappa}+\frac{p_2^\alpha p_2^\kappa}{m_{1B}^2} \bigg)p_{1\kappa}\frac{G(q^2)}{M}\frac{1}{p_1^2-m_{K}^2}\nonumber\\
&+\int_{s_1^{\rm th}}^{\infty} ds_1 \int_{s_2^{\rm th}}^{\infty} ds_2 \frac{\rho^{\rm conti}_{\mu\nu\rho}(s_1, s_2, q^2)}{(s_1-p_1^2)(s_2-p_2^2)}+\int_{s_1^{\rm th}}^{\infty} ds_1 \frac{\rho^{\rm conti}_{1,\mu\nu\rho}(s_1,p_2, q^2)}{s_1-p_1^2}+\int_{s_2^{\rm th}}^{\infty} ds_2 \frac{\rho^{\rm conti}_{2,\mu\nu\rho}(p_1, s_2, q^2)}{s_2-p_2^2}.\label{eq:hadronCorrFunc1}
\end{align}
\end{widetext}
The last three terms above denote the contribution from the excited and continuous spectrum, which begin at the thresholds $s_1^{\rm th}$ and $s_2^{\rm th}$. It should be noted that the axial  vector current $J_{\rho}^{1A}$ can create both an axial vector and a pseudoscalar  kaon from the vaccum. Therefore to obtain Eq.~\eqref{eq:hadronCorrFunc1} both the $K_{1A}$ and $K$ have been inserted between $\bar s s(0)$ and $J_{\rho}^{1A}(y)$, and we have used the parameterization for the $K\to K_{1B}$ matrix element:
\begin{eqnarray}
\langle K_{1B}(p_2)|\bar{s}s(0)|K(p_1) \rangle=\epsilon^*_\mu(p_2)p_1^\mu \frac{G(q^2)}{M},
\end{eqnarray}
where $G(q^2)$ is the corresponding form factor. 

Now the hadron level correlation function in Eq.~\eqref{eq:hadronCorrFunc1} depends on four form factors $F_{1}, F_{2}, F_{3}, G$. However, only $F_1$ is relevant to the mixing angle as shown in Eq.~\eqref{eq:thetaKformula2}. To remove the irrelevant form factors we operate the following projection on the correlation function
\begin{align}
& \epsilon^{\mu\rho\alpha\beta}p_1^\nu \Pi_{\mu\nu\rho}(p_1, p_2)={\tilde \Pi}(p_1, p_2)(p_1^{\beta}p_2^{\alpha}-p_1^{\alpha}p_2^{\beta}).\label{eq:projection}
\end{align}
${\tilde \Pi}(p_1, p_2)$ is a newly defined scalar correlation function, which at hadron level only depends on $F_1$:
\begin{align}
&{\tilde \Pi}^H(p_1, p_2)= \frac{2 m_{1A} f_{1A} f_{1B}^\perp F_1(q^2)}{(p_1^2-m_{1A}^2) (p_2^2-m_{1B}^2)}+\cdots,\label{eq:hadronCorrFunc2}
\end{align}
where the ellipse denotes the last three terms in  Eq.~\eqref{eq:hadronCorrFunc1} with the projection defined in Eq.~\eqref{eq:projection} being operated.  In principle, the ${\tilde \Pi}(p_1, p_2)$ calculated at the hadron level and the quark-gluon level should be equivalent:
\begin{align}
&{\tilde \Pi}^H(p_1, p_2,q^2)={\tilde \Pi}^{\rm QCD}(p_1, p_2,q^2)\nonumber\\
=&\frac{1}{\pi^2}\int_{0}^{\infty} ds_1 \int_{0}^{\infty} ds_2  \frac{{\rm Im}^2{\tilde \Pi}^{QCD}(s_1, s_2,q^2)}{(s_1-p_1^2)(s_2-p_2^2)}.\label{eq:BorelQCDCorr}
\end{align}
In the second equation above we have expressed the ${\tilde \Pi}^{\rm QCD}$ as its dispersive integration form, with $s_1^{\rm min}$ and $s_2^{\rm min}$ being the quark level thresholds.  According to the quark-hadron duality, the continuous spectrum contribution at hadron level is equivalent to that at QCD level. In other words the ellipse term in Eq.~\eqref{eq:hadronCorrFunc2} is equal to 
\begin{align}
\frac{1}{\pi^2}&\left[\int_{s_1^{\rm th}}^{\infty} ds_1 \int_{s_2^{\rm th}}^{\infty} ds_2+\int_{s_1^{\rm th}}^{\infty} ds_1 \int_{0}^{s_2^{\rm th}} ds_2\right.\nonumber\\
&\left.+\int_{0}^{s_1^{\rm th}} ds_1 \int_{s_2^{\rm th}}^{\infty} ds_2 \right]\frac{{\rm Im}^2{\tilde \Pi}^{\rm QCD}(s_1, s_2,q^2)}{(s_1-p_1^2)(s_2-p_2^2)}.\label{eq:BorelQCDCorr}
\end{align}
Thus the continuous spectrum contribution can be canceled at both the hadron and QCD levels. After Borel transformation we arrive at the sum rules equation:
\begin{widetext}
\begin{align}
{\cal B}_{T_1,T_2}\{{\tilde \Pi}^{H}\}(q^2)&={\cal B}_{T_1,T_2}\{{\tilde \Pi}^{\rm QCD}\}(q^2),\nonumber\\
2 m_{1A} f_{1A} f_{1B}^\perp \ e^{-\frac{m_{1A}^2}{T_1^2}} e^{-\frac{m_{1B}^2}{T_2^2} } F_1(q^2)&= \frac{1}{\pi^2}\int_{0}^{s_1^{\rm th}} ds_1 \int_{0}^{s_2^{\rm th}} ds_2\ e^{-\frac{s_1}{T_1^2}} e^{-\frac{s_2}{T_2^2} }{\rm Im}^2{\tilde \Pi}^{\rm QCD}(s_1, s_2,q^2),\label{eq:BorelHadronCorr1}
\end{align}
\end{widetext}
where $T_1, T_2$ are the two Borel parameters corresponding to $p_1^2, p_2^2$. Now it is clear that $F_1$ can be obtained through Eq.~\eqref{eq:BorelHadronCorr1} if the imaginary part of ${\tilde \Pi}^{\rm QCD}$ is calculated.

\section{Quark-gluon Level calculation}
\label{sec:QCD_level_calculation}
\subsection{Perturbative diagram}
\begin{figure*}[htp]
\centering
\includegraphics[width=0.7\textwidth]{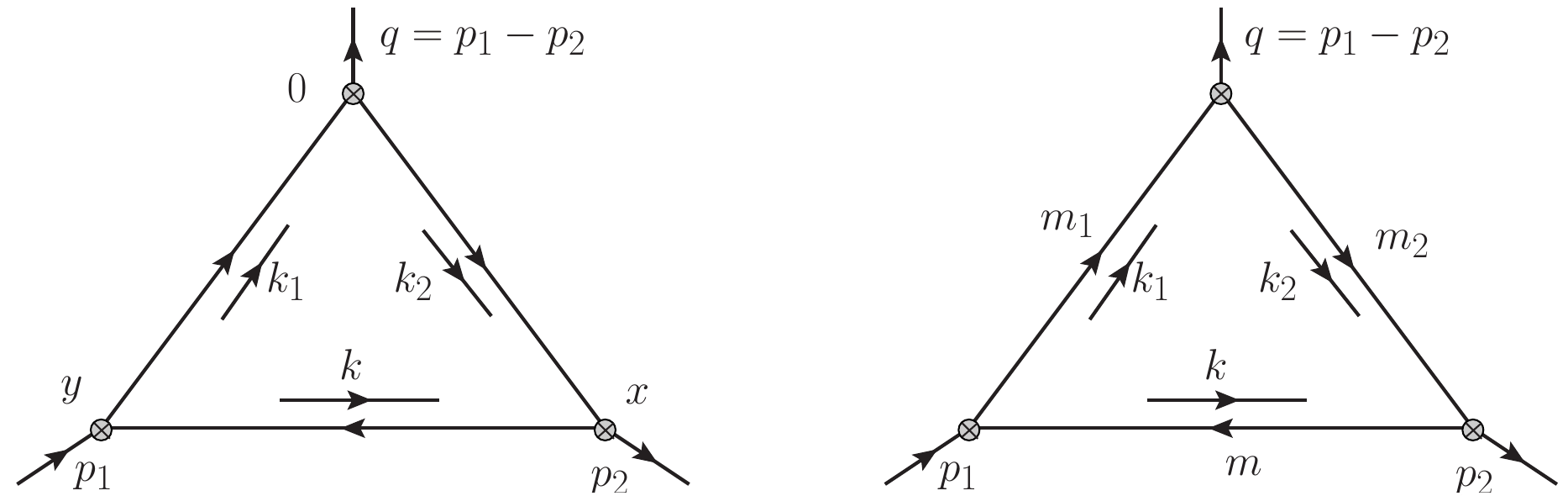}
\caption{The perturbative diagram contribution to the correlation function, where the lower two vertexes denote the kaon currents defined in Eq. \eqref{eq:kaoncurrent} (left). A general triangle diagram corresponding to a three-point correlation function (right). }
\label{fig:pert}
\end{figure*} 
In this section, we present the QCD level calculation for ${\tilde \Pi}^{\rm QCD}$ and extract its imaginary part. In the deep Euclidean region $p_1^2, p_2^2, q^2\ll 0$, ${\tilde \Pi}^{\rm QCD}$ can be analytically calculated by operator product expansion (OPE).

The leading contribution to OPE is from the perturbation diagram as shown by the left diagram in Fig.~\ref{fig:pert}, with amplitude
\begin{align}
&\Pi_{\mu\nu\rho}^{\rm pert}(p_1,p_2,q^2)\nonumber\\
=&\frac{iN_c}{(2\pi)^4}\int d^4 k_1 d^4 k_2 d^4 k \ \delta^4(p_2-k_2-k)  \delta^4(p_1-k_1-k) \nonumber\\
&\times \frac{{\rm tr}[\slashed k \sigma_{\mu\nu} (\slashed k_2+m_s)(\slashed k_1+m_s)\gamma_{\rho}\gamma_5]}{k^2 (k_2^2-m_s^2)(k_1^2-m_s^2)}.
\end{align}
The double imaginary part of the correlation function is related with its double discontinuity as
\begin{align}
&{\rm Im}^2 \Pi_{\mu\nu\rho}^{\rm pert}(p_1,p_2,q^2)= \frac{1}{(2i)^2}{\rm Disc}^2 \Pi_{\mu\nu\rho}^{\rm pert}(p_1,p_2,q^2)\nonumber\\
=& \frac{1}{(2i)^2}\frac{iN_c}{(2\pi)^4}(-2\pi i)^3 \int d\Phi_{\Delta}(p_1,p_2,m_s,m_s,0)\nonumber\\
&\times {\rm tr}[\slashed k \sigma_{\mu\nu} (\slashed k_2+m_s)(\slashed k_1+m_s)\gamma_{\rho}\gamma_5],\label{eq:pertdiagram1}
\end{align}
which is obtained by the cutting rules: ${\rm Disc} \{1/(p^2-m^2+i\epsilon)\}= (-2\pi i)\delta(p^2-m^2)$. 
In the above expression, we have introduced a three body phase space integration measure $d\Phi_{\Delta}$ for a triangle integration as shown by the right diagram in Fig.~\ref{fig:pert}:
\begin{align}
&\int d\Phi_{\Delta}(p_1,p_2,m_1,m_2,m)\nonumber\\
=&\int d^4 k_1 d^4 k_2 d^4 k\ \delta(k_1^2-m_1^2)\delta(k_2^2-m_2^2)\delta(k^2-m^2)\nonumber\\
&\times\delta^4(p_2-k_2-k)  \delta^4(p_1-k_1-k),
\end{align}
where the three internal lines are set on shell.
The scalar triangle integration with unit integrand  reads as
\begin{align}
&I_{\Delta}=\int d\Phi_{\Delta}(p_1,p_2,m_1,m_2,m)\cdot 1=\frac{\pi}{2\sqrt{\lambda}}\nonumber\\
&\times\Theta[s_1,s_2,q^2,m_1,m_2,m]\theta[s_1-s_{1}^{\rm min}]\theta[s_2-s_{2}^{\rm min}],
\end{align}
where $\lambda=(s_1+s_2-q^2)^2-4s_1 s_2$ with $p_1^2=s_1, p_2^2=s_2$. $\Theta$ is a $\theta$ function constraining the $s_1, s_2, q^2$:
\begin{align}
&\Theta[s_1,s_2,q^2,m_1,m_2,m]\nonumber\\
=&\theta\big[-m_2^4 s_1 - m_1^4 s_2 + m_2^2 [m^2 (q^2 + s_1 - s_2) \nonumber\\
 &+ s_1 (q^2 - s_1 + s_2)]\nonumber\\
 &- 
 q^2 (m^4 + s_1 s_2- m^2 (-q^2 + s_1 + s_2))\nonumber\\
 &+ 
 m_1^2 [(q^2 + s_1 - s_2) s_2\nonumber\\
 & + m^2 (q^2 - s_1 + s_2) + m_2^2 (-q^2 + s_1 + s_2)]\big].
\end{align}
The rest two $\theta$ functions ensure that $s_1, s_2$ are above the corresponding quark level thresholds , namely $s_{1,2}>s_{1,2}^{\rm min}=(m_{1,2}+m)^2$. 
In the Eq.~\eqref{eq:pertdiagram1} one have to set $m_1=m_2=m_s$. The definitions and expressions of higher rank triangle diagram integrations are given in the Appendix \ref{app:triangle}. The analytical expression of ${\rm Im}^2 {\tilde \Pi}^{\rm pert}(p_1,p_2,q^2)$ is given in the Appendix \ref{app:analyticalRes}. 

\subsection{$\bar q q$ condensate diagrams}
\begin{figure*}[htp]
\centering
\includegraphics[width=0.9\textwidth]{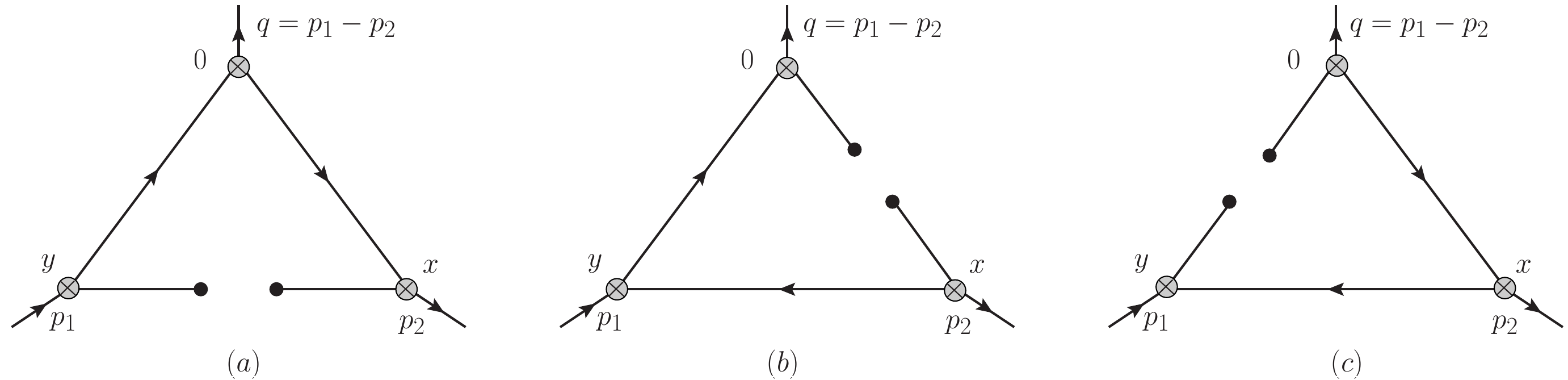}
\caption{The $\bar q q$ condensate diagram contribution to the correlation function, where one of the quark lines is disconnected. The diagrams (b) and (c) will vanish under the double Borel transformation for $p_1^2$ and $p_2^2$ simultaneously.}
\label{fig:qq}
\end{figure*}
The dimension-3 operator contribution to OPE comes from the quark condensing diagrams as shown in Fig.~\ref{fig:qq}. It can be found that the amplitudes of the $s$ quark condensing diagrams only contain one fraction form, either $1/(p_1^2-m_s^2)$ or $1/(p_2^2-m_s^2)$. Therefore, the diagrams (b) and (c) will vanish under the double Borel transformation for $p_1^2$ and $p_2^2$ simultaneously.  The amplitude of diagram (a) reads as
\begin{align} 
&\Pi_{\mu\nu\rho}^{\bar q q}(p_1,p_2,q^2)=i^2 \int d^4 x d^4 y \ e^{i p_2\cdot x} e^{-i p_1\cdot y}\nonumber\\
&\times [\sigma_{\mu\nu} D^{(0)}_s(x,0) D^{(0)}_s(0,y)\gamma_{\rho}\gamma_5] \langle 0|{\bar q}_a^i(x) q_b^i(y)|0\rangle,\label{eq:qbarqCorr1}
\end{align}
where $D^{(0)}_s$ denotes free $s$ quark propagator. The nonlocal $\bar q q$ condensing matrix element can be expanded up to dimension-5 local operators as 
\begin{align}
 &\langle 0|{\bar q}_a^i(x) q_b^i(y)|0\rangle\nonumber\\
 =& N_c\left[\langle\bar q q\rangle \frac{1}{12}\delta_{ba}+\langle\bar q G q\rangle  \frac{1}{192}(x-y)^2\delta_{ba}\right].\label{eq:nonlocalqq}
\end{align}
The contribution of dimension-3 operator, namely the  $\bar q q$ condensate, only comes from the first term given above. The  corresponding amplitude reads as
\begin{align}
&{\rm Im}^2{\tilde \Pi}^{\bar q q}(p_1,p_2,q^2)\nonumber\\
=&\frac{2\pi^2}{3}N_c m_s \langle\bar q q\rangle\delta(s_1-m_s^2)\delta(s_2-m_s^2).
\end{align}

The second term in Eq.~\eqref{eq:nonlocalqq} provides a contribution from the dimension-5 operator $\bar q g_s G_{\alpha\beta} q$. The corresponding amplitude reads as
\begin{align}
&\Pi_{\mu\nu\rho}^{\bar q G q(1)}(p_1,p_2,q^2)\nonumber\\
=&\frac{N_c}{192}\langle\bar q G q\rangle(-1)\left(\frac{\partial^2}{\partial p_2^{\alpha}\partial p_{2\alpha}}+\frac{\partial^2}{\partial p_1^{\alpha}\partial p_{1\alpha}}+2\frac{\partial^2}{\partial p_1^{\alpha}\partial p_{2\alpha}}\right)\nonumber\\
&\times {\rm tr}[\sigma_{\mu\nu}(\slashed p_2+m_s)(\slashed p_1+m_s)\gamma_{\rho}\gamma_5]\nonumber\\
&\times\frac{1}{p_1^2-m_s^2}\frac{1}{p_2^2-m_s^2},\label{eq:qbarqCorr2}
\end{align}
where we have transformed $x, y$ to $-i\partial/\partial p_2, i\partial/\partial p_1$ through the exponential terms in Eq.~\eqref{eq:qbarqCorr1}. To simplify the calculation of Eq.~\eqref{eq:qbarqCorr2} we can omit the terms suppressed by $m_s^2$ and obtain
\begin{align}
&\Pi_{\mu\nu\rho}^{\bar q G q(1)}(p_1,p_2,q^2)\nonumber\\
=&-\frac{N_c}{12}\langle\bar q G q\rangle m_s  \epsilon_{\mu\nu\rho\alpha}(p_1^{\alpha}-p_2^{\alpha})\frac{\partial}{\partial M^2}\nonumber\\
&\times\left[\frac{1}{(p_1^2-M^2)(p_2^2-m_s^2)}-\frac{1}{(p_1^2-m_s^2)(p_2^2-M^2)}\right]|_{M^2=m_s^2}\nonumber\\
-&\frac{N_c}{12}\langle\bar q G q\rangle m_s  \epsilon_{\mu\nu\rho\alpha}(p_1^{\alpha}+p_2^{\alpha}) q^2 \nonumber\\
&\times\frac{\partial^2}{\partial M_1^2 \partial M_2^2}\left[\frac{1}{(p_1^2-M_1^2)(p_2^2-M_2^2)}\right]|_{M_1^2=M_2^2=m_s^2}.
\end{align}
To obtain the above expression, we have introduced derivatives on  auxiliary masses $M_1,M_2$ to lower the power of the denominator, which means
\begin{align}
    \frac{1}{(p_{1,2}^2-m_{1,2}^2)^2}=\frac{\partial}{\partial M^2}\frac{1}{p_{1,2}^2-M^2}|_{M^2\to m_{1,2}^2}.
\end{align}
Taking the imaginary part, using dispersive integration and conducting Borel transformation, we arrives at
\begin{align}
&{\cal B}_{T_1,T_2}\{{\tilde \Pi}^{\bar q G q(1)}\}(q^2)=\frac{1}{6}N_c m_s\langle\bar q G q\rangle\nonumber\\
&\times e^{-m_s^2/T_1^2-m_s^2/T_2^2}\left[\frac{1}{T_1^2}-\frac{1}{T_2^2}+\frac{q^2}{T_1^2 T_2^2}\right].
\end{align}

\subsection{$\bar q G q$ condensate diagrams}
\begin{figure*}[htp]
\centering
\includegraphics[width=0.9\textwidth]{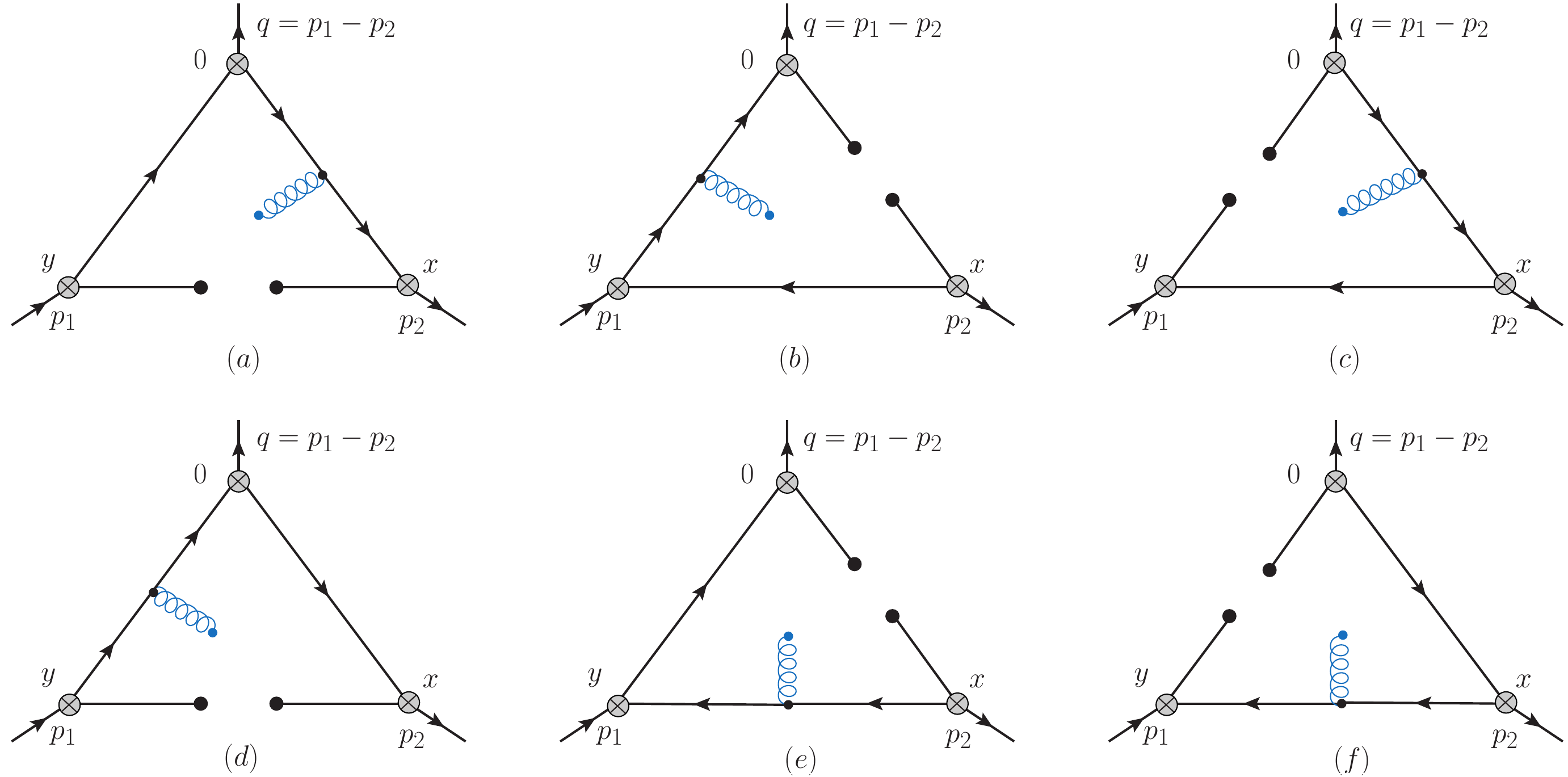}
\caption{The $\bar q G q$ condensate diagram contribution to the correlation function, where one of the quark line emits a solft gluon which condensing with other two disconnected quark fields. The diagrams (b), (c), (e) and (f) vanishes under double Borel transformation. The diagram (d) vanishes under the projection introduced in Eq.~\eqref{eq:projection}.}
\label{fig:qGq}
\end{figure*}

The quark-gluon condensing diagrams as shown in Fig.~\ref{fig:qGq}, where a quark interacts with a background gluon field which condensates with the other two disconnected light quark fields. These diagrams provide the  dimension-5 operator contribution in the OPE. The massive and massless  quark propagators in the background gluon field read as
\begin{widetext}
\begin{align}
D_s(x,0)=&\ i\int \frac{d^4 k}{(2\pi)^4}e^{-i k\cdot x}\left[\frac{\delta_{ij}}{\slashed k -m_s}-\frac{g_s G_{\alpha\beta}^A t_{ij}^A}{4}\frac{\sigma^{\alpha\beta}(\slashed k+m_s)+(\slashed k+m_s)\sigma^{\alpha\beta}}{(k^2-m_s^2)^2}\right.\nonumber\\
&\left.-\frac{g_s^2(t^A t^B)_{ij}G_{\alpha\beta}^A G_{\mu\nu}^B[f^{\alpha\beta\mu\nu}(k)+f^{\alpha\mu\beta\nu}(k)+f^{\alpha\mu\nu\beta}(k)]}{4(k^2-m_s^2)^2}+\cdots\right],\nonumber\\
D_q(x,0)=&\ \frac{i\delta_{ij}\slashed x}{2\pi^2 x^4}-\frac{i g_s G_{\alpha\beta}^A t_{ij}^A (\slashed x \sigma^{\alpha\beta} + \sigma^{\alpha\beta} \slashed x)}{32 \pi^2 x^2}+\cdots,\label{eq:propagators}
\end{align}
where
\begin{align}
    f^{\alpha\beta\mu\nu}(k)=\  (\slashed k +m_s)\gamma^{\alpha}(\slashed k +m_s)\gamma^{\beta}(\slashed k +m_s)\gamma^{\mu}(\slashed k +m_s)\gamma^{\nu}(\slashed k +m_s),
\end{align}

\end{widetext}
and we have only present the terms relevant to the OPE up to dimension-5 operators in Eq.\eqref{eq:propagators}.

It can be found that the diagram (b), (c), (e) and (f) in Fig.~\ref{fig:qGq} vanish after the Borel transformation due to the same reason as that happens in the $\bar q q$ condensate diagrams. The diagram (d) vanishes after the projection introduced in Eq.~\eqref{eq:projection}. Thus diagram (a) is the only non-vanishing diagram with amplitude:
\begin{widetext}
\begin{align}
\Pi_{\mu\nu\rho}^{\bar q G q (2)}(p_1,p_2,q^2)
= &\  i^2 \int d^4 x d^4 y \ e^{i p_2\cdot x} e^{-i p_1\cdot y}\int\frac{d^4 k_1}{(2\pi)^4}\frac{d^4 k_2}{(2\pi)^4}\ e^{i k_1\cdot y} e^{-i k_2\cdot x} \nonumber\\
&\times \left[\sigma_{\mu\nu}\left(-i\over 4\right)\frac{\sigma^{\alpha\beta}(\slashed k_2 +m_s)+(\slashed k_2 +m_s)\sigma^{\alpha\beta}}{(k_2^2-m_s^2)^2}\frac{i(\slashed k_1 +m_s)}{k_1^2-m_s^2}\gamma_{\rho}\gamma_5\right]_{ab}\nonumber\\
&\times t^{A}_{ij} \langle 0|{\bar q}_a^i(x) g_s G_{\alpha\beta}^A(0)q_b^i(y)|0\rangle.
\end{align}
\end{widetext}
Using the the quark-gluon condensate formula, keeping the leading term: 
\begin{align}
\langle 0|{\bar q}_a^i(x) g_s G_{\alpha\beta}^A(0)q_b^i(y)|0\rangle=\frac{1}{192}\langle \bar q G q\rangle (\sigma_{\alpha\beta})_{ba} t_{ji}^A+\cdots,
\end{align}
and conducting the projection introduced in Eq.~\eqref{eq:projection}, we can obtain 
\begin{align}
&{\tilde \Pi}^{\bar q G q (2)}(p_1,p_2,q^2)\nonumber\\
=&\frac{1}{24}m_s\langle \bar q G q\rangle \frac{\partial}{\partial M^2}\frac{1}{(p_1^2-m_s^2)(p_2^2-M^2)}|_{M^2=m_s^2}.
\end{align}
The Borel transformed form reads as
\begin{align}
&{\cal B}_{T_1,T_2}\{{\tilde \Pi}^{\bar q G q(2)}\}(q^2)\nonumber\\
=&-\frac{1}{6} m_s \langle \bar q G q\rangle \frac{1}{T_2^2} e^{-m_s^2/T_1^2}e^{-m_s^2/T_2^2}.
\end{align}

\subsection{GG condensate diagrams}
\begin{figure*}[htp]
\centering
\includegraphics[width=0.9\textwidth]{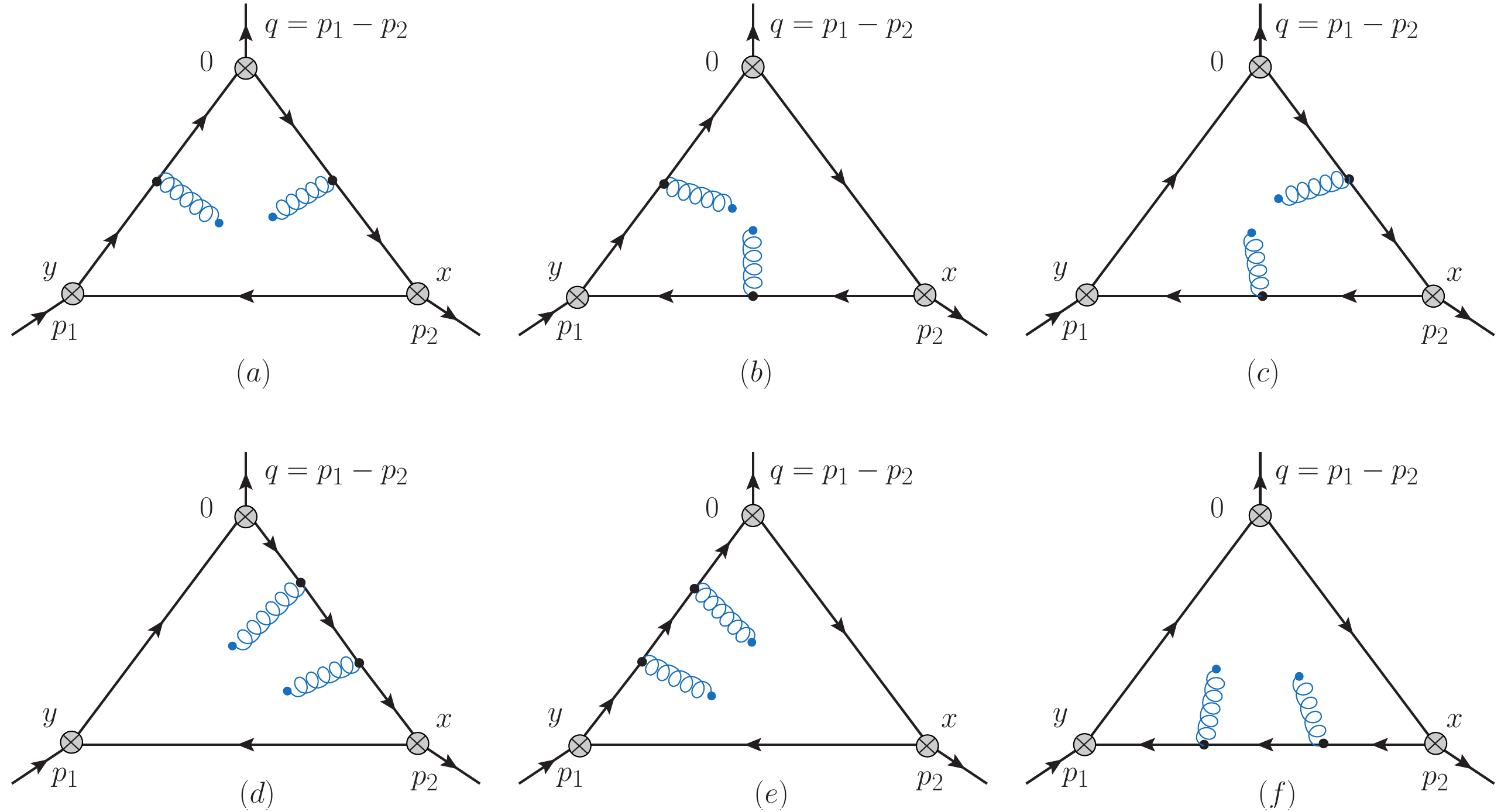}
\caption{The $GG$ condensate diagram contribution to the correlation function, where two soft gluons are emitted from the internal quark lines and condensate with each other. The diagrams (d), (e) and (f) can be neglected in this calculation since the diagrams (d), (e) are suppressed by $m_s^2$, while the diagram (f) is suppressed by $m^2$.}
\label{fig:GG}
\end{figure*}
The gluon-gluon condensate diagrams are shown in Fig.~\ref{fig:GG}, where the internal quarks interact with two soft background gluon fields which condensate in the vaccum. These diagrams provide the  dimension-4 operator contribution in the OPE. In the diagrams (a), both the two $s$ 
quark interact with the background gluons, and the corresponding amplitude reads as
\begin{widetext}
\begin{align}
&\Pi_{\mu\nu\rho}^{GG(a)}(p_1,p_2,q^2)\nonumber\\
=&\int d^4 x d^4 y \ e^{i p_2\cdot x} e^{-i p_1\cdot y}\int\frac{d^4 k_1}{(2\pi)^4}\frac{d^4 k_2}{(2\pi)^4}\frac{d^4 k}{(2\pi)^4}\ e^{i k_1\cdot y} e^{-i k_2\cdot x} e^{-i k\cdot (y-x)}\left(-\frac{i}{4}\right)^2\nonumber\\
&\times {\rm tr}\left[\frac{-i\slashed k}{k^2}\sigma_{\mu\nu} \frac{\sigma^{\alpha\beta}(\slashed k_2 +m_s)+(\slashed k_2 +m_s)\sigma^{\alpha\beta}}{(k_2^2-m_s^2)^2}\frac{\sigma^{\kappa\tau}(\slashed k_1+m_s)+(\slashed k_1 +m_s)\sigma^{\kappa\tau}}{(k_2^2-m_s^2)^2}\gamma_{\rho}\gamma_5\right]\nonumber\\
&\times {\rm tr}[t^A t^A] g_s^2 \langle 0 | G_{\alpha\beta}^A(0) G_{\kappa\tau}^A(0) | 0\rangle.
\end{align}
\end{widetext}
Using the gluon condensate formula:
\begin{align}
& g_s^2 \langle 0 | G_{\alpha\beta}^A(0) G_{\kappa\tau}^A(0) | 0\rangle\nonumber\\
=&\frac{1}{96}\langle GG \rangle \delta_{AB} (g_{\alpha\kappa}g_{\beta\tau}-g_{\alpha\tau}{\beta\kappa}),
\end{align}
and extracting the imaginary part by cutting rules, we arrive at
\begin{align}
&{\rm Im}^2\Pi_{\mu\nu\rho}^{GG(a)}(p_1,p_2,q^2)\nonumber\\
=&\frac{1}{3072 \pi}  \frac{\partial^2}{\partial M_1^2\partial M_2^2}\int d\Phi_{\Delta}[p_1,p_2,M_1,M_2,0]\nonumber\\
&\times {\rm tr}\left[\slashed k\sigma_{\mu\nu}(\sigma^{\alpha\beta}(\slashed k_2 +m_s)+(\slashed k_2 +m_s)\sigma^{\alpha\beta})\right.\\
&\left.\times(\sigma^{\kappa\tau}(\slashed k_1+m_s)+(\slashed k_1 +m_s)\sigma^{\kappa\tau})\gamma_{\rho}\gamma_5\right]\nonumber\\
&\times (g_{\alpha\kappa}g_{\beta\tau}-g_{\alpha\tau}{\beta\kappa})|_{M_1^2=m_s^2,M_2^2 =m_s^2}.
\end{align}
Note that before doing the derivative on the auxiliary masses we must temporary change the invariant mass square of the $k_1, k_2$ lines to be $M_1^2, M_2^2$. The analytical result of ${\rm Im}^2\tilde\Pi^{GG(k_1k_2)}$ is given in the Appendix \ref{app:analyticalRes}.

In the diagrams (b) and (c), one of the two condensing gluons comes from the massless quark. The massless quark propagator in the background gluon field has been  given by Eq.~\eqref{eq:propagators} with the use of coordinate space. The amplitude of diagram (b) reads as
\begin{widetext}
\begin{align}
&\Pi_{\mu\nu\rho}^{GG(b)}(p_1,p_2,q^2)\nonumber\\
=&\int d^4 x d^4 y \ e^{i p_2\cdot x} e^{-i p_1\cdot y}\int\frac{d^4 k_1}{(2\pi)^4}\frac{d^4 k_2}{(2\pi)^4}\ e^{i k_1\cdot y} e^{-i k_2\cdot x}\left(-\frac{i}{32 \pi^2}\right)\left(-\frac{i}{4}\right) {\rm tr}[t^A t^B] \nonumber\\
&\times {\rm tr}\left[\frac{(\slashed y- \slashed x)\sigma^{\kappa\tau}+\sigma^{\kappa\tau}(\slashed y- \slashed x)}{(y-x)^2}\sigma_{\mu\nu} \frac{i(\slashed k_2 +m_s)}{k_2^2-m_s^2}\frac{\sigma^{\alpha\beta}(\slashed k_1 +m_s)+(\slashed k_1 +m_s)\sigma^{\alpha\beta}}{(k_1^2-m_s^2)^2}\gamma_{\rho}\gamma_5\right]\nonumber\\
&\times  g_s^2 \langle 0 | G_{\alpha\beta}^A(0) G_{\kappa\tau}^B(0) | 0\rangle.
\end{align}
\end{widetext}
Redefining the coordinate: $w=y-x$, 
and using the integration formula in the coordinate space:
\begin{align}
\int d^4 w\  e^{-i(p_1-k_1)\cdot w}\frac{1}{w^2}=(-4\pi^2 i)\frac{1}{(p_1-k_1)^2},
\end{align}
we have the imaginary part as
\begin{align}
&\ {\rm Im}^2 {\tilde \Pi}^{GG(b)}(p_1,p_2,q^2)(p_1^{\beta^{\prime}}p_2^{\alpha^{\prime}}-p_2^{\alpha^{\prime}}p_2^{\beta^{\prime}}) \nonumber\\
=& -\frac{1}{6144 \pi} \langle GG \rangle\epsilon^{\mu\rho\alpha^{\prime}\beta^{\prime}}p_1^{\nu}\frac{\partial}{\partial M_1^2}\left(\frac{\partial}{\partial p_1^{\sigma}}+\frac{\partial}{\partial p_2^{\sigma}}\right)\nonumber\\
&\times \int d\Phi_{\Delta}[p_1,p_2,M_1,m_s,0] [g_{\alpha\kappa}g_{\beta\tau}-g_{\alpha\tau}g_{\beta\kappa}]\nonumber\\
&\times {\rm tr}[(\gamma^{\sigma}\sigma^{\kappa\tau}+\sigma^{\kappa\tau}\gamma^{\sigma})\sigma_{\mu\nu}(\sigma^{\alpha\beta}(\slashed k_2 +m_s)\nonumber\\
&+(\slashed k_2 +m_s)\sigma^{\alpha\beta})(k_1^2-m_s^2)\gamma_{\rho}\gamma_5]|_{M_1^2=m_s^2},
\end{align}
where the linear term of $w$ has been transformed to the derivatives of $p_1, p_2$. The calculation of diagram (c) is almost the same so it will not be present here. On the other hand, it can be found that the diagrams (d), (e) and (f) can be neglected in this calculation, since the diagrams (d), (e) are suppressed by $m_s^2$, while the diagram (f) is suppressed by $m^2$. Therefore these three diagrams will not be considered here. The analytical results for the diagrams (a), (b) and (c) are given in the Appendix \ref{app:analyticalRes}.

\section{Numerical Results}
\label{sec:Numerical}
The masses of $K_{1A}, K_{1B}$ and their decay constants are taken from Ref.\cite{Yang:2007zt}: $m_{K_{1A}}=1.31\pm 0.06\ {\rm GeV}$, $m_{K_{1B}}=1.34\pm 0.08\ {\rm GeV}$, $f_{1A}=0.25\pm 0.013\ {\rm GeV}$ and $f_{1B}=0.19\pm 0.01\ {\rm GeV}$. In this work we set $m_u=m_d=0$, and $m_s=(0.1\pm 0.005)\ {\rm GeV}$ at the energy scale $\mu=m_{K_1}=1.3\  {\rm GeV}$ \cite{ParticleDataGroup:2020ssz}. The condensate parameters are taken as \cite{Ioffe:2005ym,Colangelo:2000dp}: $\langle\bar{q} q\rangle=-(0.24 \pm 0.01 \mathrm{GeV})^3,\left\langle\bar{q} G q\right\rangle=m_0^2\langle\bar{q} q\rangle \ {\rm with}\  m_0^2=(0.8 \pm 0.2) \mathrm{GeV}^2, \langle G G\rangle=(4\pi^2)(0.012 \pm 0.004) \mathrm{GeV}^4$.

In terms of the threshold parameters, note that since $J^{1A}_{\rho}$ can create both pseudoscalar and axial vector kaons, $s_1^{\rm th}$ should correspond to the excited pseudoscalar $K(1460)$. On the other hand, $J^{1B}_{\mu\nu}$ can only create axial vector kaons, thus $s_2^{\rm th}$ should correspond to the excited axial vector $K_1(1650)$. Therefore we choose $s_1^{\rm th}=1.46^2\ {\rm GeV}^2 $ and $s_2^{\rm th}=1.65^2\ {\rm GeV}^2 $.

\begin{figure}[htp]
\centering
\includegraphics[width=1\textwidth]{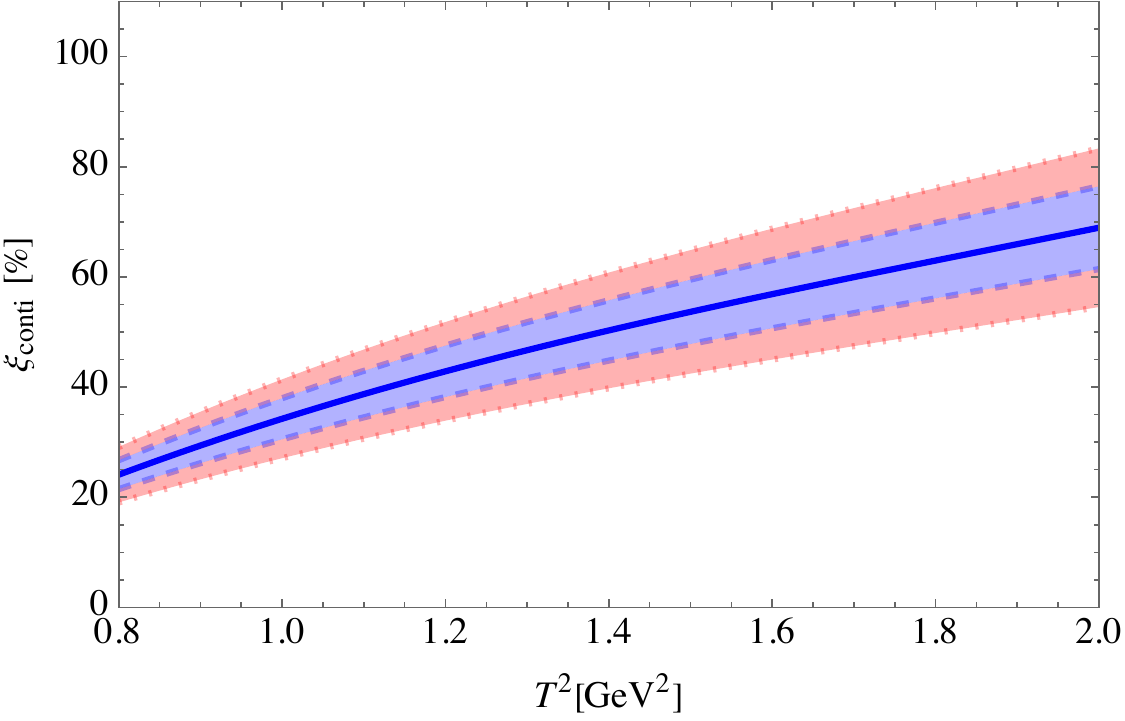}
\caption{$\xi_{\rm conti}$ as a function of $T^2$  with $q^2=-6\ {\rm GeV}^2$. The blue and red bands denote the errors from the uncertainties of condensate parameters and $m_{K_{1A}},\ f_{1A},\ f_{1B}^{\perp}$, respectively.}
\label{fig:xicontietacond}
\end{figure}

The determination of Borel parameters depends on two criteria. Firstly, the contribution from the continuous spectrum must be suppressed so that be smaller than the pole contribution. Quantitatively, this criterion can be expressed by the constraint:
\begin{align}
\xi_{\rm conti}&\equiv\frac{\int_{s_1^{\rm th}}^{\infty} ds_1 \int_{s_2^{\rm th}}^{\infty} ds_2\ e^{-\frac{s_1}{T_1^2}} e^{-\frac{s_2}{T_2^2} }{\rm Im}^2{\tilde \Pi}^{\rm QCD}(s_1, s_2,q^2)}{\int_{0}^{\infty} ds_1 \int_{0}^{\infty} ds_2\ e^{-\frac{s_1}{T_1^2}} e^{-\frac{s_2}{T_2^2} }{\rm Im}^2{\tilde \Pi}^{\rm QCD}(s_1, s_2,q^2)}\nonumber\\
&\lesssim 0.5,\label{eq:BorelHadronCorr}
\end{align}
where the nominator denotes the contribution from the continuous spectrum, while the denominator denotes all the spectrum contribution. Note that the Borel parameter is related with the corresponding hadron mass. Since the mass difference of the intial and final kaon is little, one can simply set $T_1=T_2=T$. Without loss of generality, $q^2$ can be chosen by an arbitrary value in the deep Euclidean region when determining $T^2$. Here we choose $q^2=-6\  {\rm GeV}^2$ and present $\xi_{\rm conti}$ as a function of $T^2$  in Fig.\ref{fig:xicontietacond},  where the blue and red bands denote the errors from the uncertainties of condensate parameters and $m_{K_{1A}},\ f_{1A},\ f_{1B}^{\perp}$, respectively.
It can be found that $\xi_{\rm conti}$ increases with the increasing of $T^2$. $\xi_{\rm conti}=50 \% $ 
occurs at $T^2=1.16 \thicksim 1.81\ {\rm GeV}^2$, which gives the range of the upper limit  for the Borel parameter: $1.16\ {\rm GeV}^2 < T_{\rm upper}^2<1.81\ {\rm GeV}^2$.

\begin{figure*}[htp]
\centering
\includegraphics[width=0.48\textwidth]{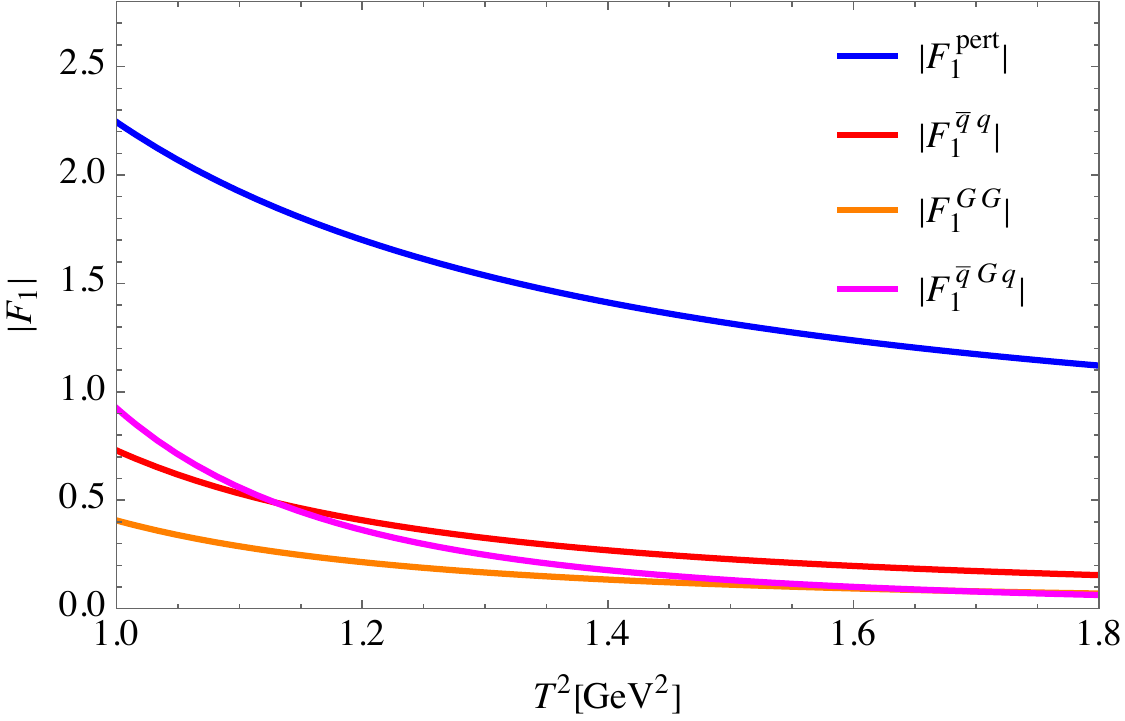}
\includegraphics[width=0.47\textwidth]{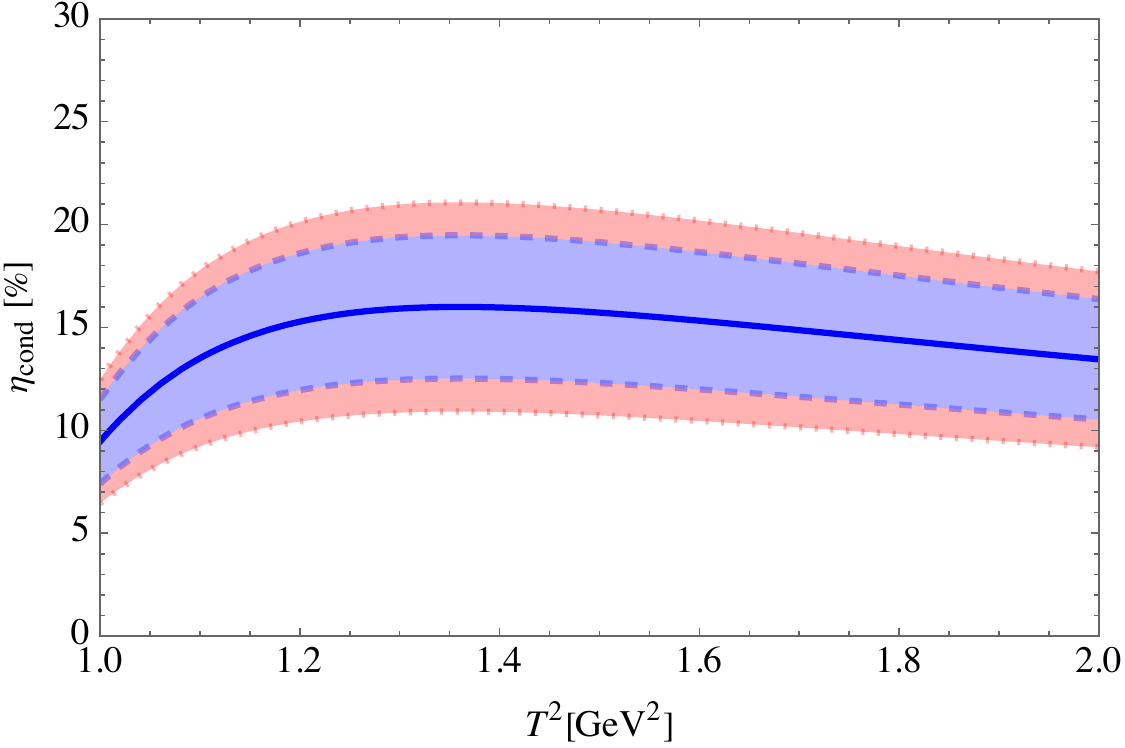}
\caption{Absolute $F_1$ contributed from various of condensate diagrams (left) and $\eta_{\rm cond}$ as a function of $T^2$ (right) with $q^2=-6\ {\rm GeV}^2$. The blue and red bands denote the errors from the uncertainties of condensate parameters and $m_{K_{1A}},\ f_{1A},\ f_{1B}^{\perp}$, respectively.}
\label{fig:etacond}
\end{figure*}
The second criterion demands the convergence of OPE, which means that the $F_1$ contributed from the higher dimension condensates must be smaller than that from the lower ones, namely:
\begin{align}
|F_1^{\rm pert}|>|F_1^{\bar q q}|>|F_1^{GG}|>|F_1^{\bar q G q}|.\label{eq:conditionOPE1}
\end{align}
The left diagram in Fig.\ref{fig:etacond} shows the absolute $F_1$ contributed from various of condensate diagrams with $q^2=-6\ {\rm GeV}^2$. It is obvious that the perturbative diagram contribution is larger than all the other condensate contributions. The right diagram in Fig.\ref{fig:etacond} shows the fraction of the condensate and the perturbative contribution: 
\begin{align}
\eta_{\rm cond}\equiv\frac{|F_1^{\bar q q}+F_1^{GG}+F_1^{\bar q G q}|}{|F_1^{\rm pert}|}.
\end{align}
It can be found that $\eta_{\rm cond}<1$ is safely satisfied in a wide $T^2$ range. Furthermore, the comparison between the condensate contribution themselves leads to
\begin{align}
|F_1^{\bar q q}|>|F_1^{GG}+F_1^{\bar q G q}| &\ \Longrightarrow\  T^2 \gtrsim 1.58\ {\rm GeV}^2, \nonumber\\
 |F_1^{GG}|>|F_1^{\bar q G q}| &\ \Longrightarrow\  T^2 \gtrsim 1.68\ {\rm GeV}^2.
\end{align}
Note that the critical points obtained above are within the upper limit range given by the first criterion $1.16\ {\rm GeV}^2 < T_{\rm upper}^2<1.81\ {\rm GeV}^2$, which determines a narrow window for $T^2$: $1.6\ {\rm GeV}^2 \lesssim T^2 \lesssim 1.8\ {\rm GeV}^2$.

\begin{figure}[h]
\centering
\includegraphics[width=1.0\textwidth]{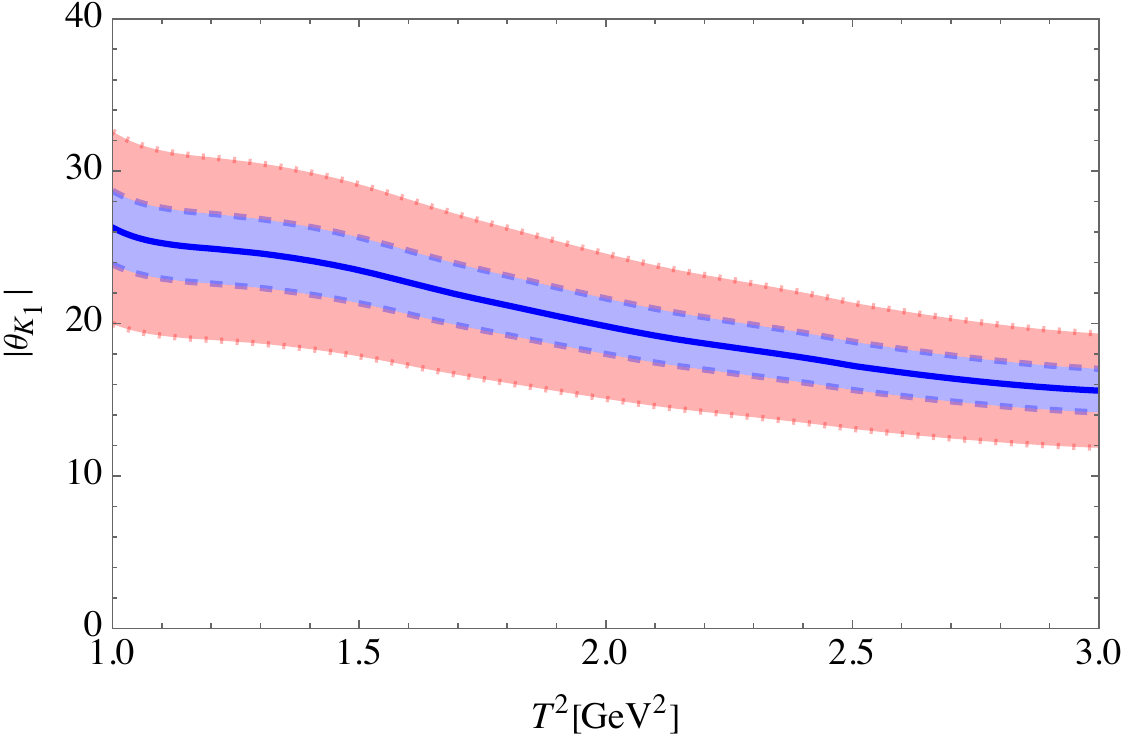}
\caption{Absolute value of $\theta_{K_1}$ as a function of $T^2$. The $\theta_{K_1}$ is calculated from Eq.\eqref{eq:thetaKformula2}, with $F_1(0)$ being fitted by Eq. \eqref{eq:fitfunction} in the region $-10\  {\rm GeV}^2< q^2 <-3\  {\rm GeV}^2$. The blue and red bands denote the errors from the uncertainties of condensate parameters and $m_{K_{1A}},\ f_{1A},\ f_{1B}^{\perp}$, respectively.}
\label{fig:thetaK}
\end{figure}
Finally, we have to check the behavior  of $F_1(0)$ in the Borel parameter window $1.6\ {\rm GeV}^2 \lesssim T^2 \lesssim 1.8\ {\rm GeV}^2$. The physical region result: $F_1(q^2\geq0)$ can be obtained by using  the deep Euclidean values $F_1(q^2\leq 0)$ and analytic continuation.  In this work, a single pole formula is used:
\begin{align}
    F_1 (q^2)=\frac{F_1(0)}{1-q^2/m_{\rm pole}^2}\label{eq:fitfunction}
\end{align}
to fit 
$F_1(q^2)$ in a region $-10\  {\rm GeV}^2< q^2 <-3\  {\rm GeV}^2$, which is chosen so that the spectral integrals can be calculated safely by applying cutting rules. In Eq.\eqref{eq:fitfunction}, $F_1(0)$ and $m_{\rm pole}$ play the role of fitting parameters, and the mixing angle $\theta_{K_1}$ can be obtained from $F_1(0)$ through Eq. \eqref{eq:thetaKformula2}. It should be mentioned that the decay constants $f_{1A}$ and $f_{1B}^{\perp}$ in Eq. \eqref{eq:BorelHadronCorr1} are calculated by QCDSR in Ref.\cite{Yang:2007zt}, which have sign ambiguity. The reason is that when using a two-point correlation in QCDSR to calculate  $f_{1A}$ or $f_{1B}^{\perp}$, one can only determine their square and thus the exact sign cannot be determined. Considering the sign ambiguity, we present the absolute value of $\theta_{K_1}$ as a function of $T^2$ in Fig.\ref{fig:thetaK}. 
Including the effect of error bands,  we obtain the  mixing angle as: $|\theta_{K_1}|=22^{\circ}\pm 7^{\circ}$. Note that both $\theta_{K_1}$ and $90^{\circ}-\theta_{K_1}$ are the solutions to Eq. \eqref{eq:thetaKformula2}. Therefore another possible mixing angle value is: $|\theta_{K_1}^{\prime}|=68^{\circ}\pm 7^{\circ}$.

\begin{table}[h]
  \caption{Comparing the $|\theta_{K_1}|$ obtained in this work with those from literature. }
\label{Tab:comparethetaK}
\begin{tabular}{|c|c|}
\hline
\hline
Refs. & $|\theta_{K_1}|$  \tabularnewline
\hline 
This Work & $22^{\circ}\pm 7^{\circ}$ or $68^{\circ}\pm 7^{\circ}$  \tabularnewline
(1) \cite{Suzuki:1993yc} & $33^{\circ}$ or  $57^{\circ}$\tabularnewline
(2) \cite{Burakovsky:1997dd} & $34^{\circ}<|\theta_{K_1}|<55^{\circ}$ \tabularnewline
(3) \cite{Cheng:2003bn,Li:2009tx} & $37^{\circ}$ or  $58^{\circ}$ \tabularnewline
(4) \cite{Close:1997nm} & $\left(31.7_{-2.5}^{+2.8}\right)^{\circ}$ or $\left(56.3_{-4.1}^{+3.9}\right)^{\circ}$ \tabularnewline
(5) \cite{Cheng:2013cwa} & $28^{\circ}<|\theta_{K_1}|<30^{\circ}$ \tabularnewline
(6) \cite{Dag:2012zz} & $39^{\circ}\pm 4^{\circ}$ \tabularnewline
\hline 
\end{tabular}
\end{table}
In Table \ref{Tab:comparethetaK}, we compare our result on $|\theta_{K_1}|$ with those obtained in literature by various of methods:
\begin{enumerate}[(1)]
\item Using early experimental information on masses and the partial rates of $K_1(1270)$ and $K_1(1400)$ \cite{Suzuki:1993yc};
\item Using non-relativistic constituent quark model with the inputs of the mass difference between the $a_1(1260)$ and $b_1(1235)$ mesons, as well as the ratio of the constituent quark masses \cite{Burakovsky:1997dd};
\item Phenomenologically analyzing on the $\tau$ weak decays: $\tau \to K_1(1270)\nu_{\tau}$ and $\tau \to K_1(1400)\nu_{\tau}$ \cite{Cheng:2003bn,Li:2009tx};
\item Analyzing the $f_1(1285)-f_1(1420)$ mixing angle $\theta_{^3 P_1}$ and its correlation to $\theta_{K_1}$ \cite{Close:1997nm};
\item Analyzing both the mixing angle of $f_1(1285)-f_1(1420)$ and $h_1(1170)-h_1(1380)$ \cite{Cheng:2013cwa};
\item Relating $\theta_{K_1}$ with a two-point correlation function, which is studied by QCDSR.\cite{Dag:2012zz}.
\end{enumerate}
It can be found that most of the $|\theta_{K_1}|$ value in the literature are in the vicinity of either $33^{\circ}$ or $57^{\circ}$. Our result:  $|\theta_{K_1}|=22^{\circ}\pm 7^{\circ}$ is slightly below this range but consistent with that given by Ref.  \cite{Cheng:2013cwa} within the error. In Ref.  \cite{Cheng:2013cwa}, to determine $\theta_{K_1}$ the authors found the correspondence between $\theta_{K_1}$ and  the  $f_1(1285)-f_1(1420)$, $h_1(1170)-h_1(1380)$ mixing angles, and ruled out unreasonable $\theta_{K_1}$ values in previous literature. In Ref. \cite{Dag:2012zz}, a different QCDSR program was performed to extract $\theta_{K_1}$, where the authors related $\theta_{K_1}$ with a two-point correlation function (Eq.[2] in Ref. \cite{Dag:2012zz}) and calculated it by OPE. However, when introducing the interpolation current of $K_{1A}$, the authors missed the contribution from the pseudoscalar $K$ as illustrated in section \ref{sec:hadron_level_calculation}, and wrongly extracted the longitudinal component of the two-point correlation function.

\section{summary}
\label{sec:summary}
In this work we investigate the $K_1(1270)-K_1(1400)$ mixing caused by the flavor $SU(3)$ symmetry breaking. The mixing angle is expressed by a $K_{1A}\to K_{1B}$ matrix element induced by the $s$ quark mass operator that breaks flavor $SU(3)$ symmetry. We focus on the QCD contribution to this matrix element and calculate it by QCDSR, where a three-point correlation function is defined and calculated both at the hadron and quark-gluon levels. In the calculation at the quark-gluon level, the operator product expansion is up to dimension-5 condensates. A detailed numerical analysis is performed to determine the Borel parameters, and the obtained mixing angle is $\theta_{K_1}=22^{\circ}\pm 7^{\circ}$ or $\theta_{K_1}=68^{\circ}\pm 7^{\circ}$, which is consistent with the phenomenological analysis on the relation between $\theta_{K_1}$ and the mixing angle of strangeless axial vector mesons. 

\section*{Acknowledgement}

We thank Wei Wang and Zhen-Xing Zhao for valuable discussions. This work is supported in part by Natural Science Foundation of China under Grant No.12305103,
12205180, and 12147140. The work of Y.J. Shi is also supported by Opening Foundation of Shanghai Key Laboratory of Particle Physics and Cosmology under Grant No.22DZ2229013-2. The work of J.Zeng is also partially supported by the Project funded by China Postdoctoral Science Foundation under Grant No. 2022M712088.

\begin{appendix}
\section{Triangle Diagram Integration}\label{app:triangle}

Here we present the rank one and two triangle diagram integrations. They are defined as
\begin{widetext}
\begin{align}
\int d\Phi_{\Delta}(p_1,p_2,m_1,m_2,m) k^{\mu}&=(A_1 p_1^{\mu}+B_1 p_2^{\mu})I_{\Delta},\nonumber\\
\int d\Phi_{\Delta}(p_1,p_2,m_1,m_2,m)\ k^{\mu_1}k^{\mu_2}&=[A_2 p_1^{\mu_1}p_1^{\mu_2}+B_2 p_2^{\mu_1}p_2^{\mu_2}+C_2 (p_1^{\mu_1}p_2^{\mu_2}+p_2^{\mu_1}p_1^{\mu_2})+D_2 g^{\mu_1\mu_2}]I_{\Delta},
\end{align}
where
\begin{align}
A_1=&\frac{-\left(m^2 (q^2-s_1+s_2)\right)+s_2
   \left(2 m_1^2-q^2-s_1+s_2\right)+m_2^2
   (q^2-s_1-s_2)}{q^4-2 q^2
   (s_1+s_2)+(s_1-s_2)^2},\nonumber\\
B_1=&\frac{-\left(m^2 (q^2+s_1-s_2)\right)+m_1^2
   (q^2-s_1-s_2)+s_1 \left(2
   m_2^2-q^2+s_1-s_2\right)}{q^4-2
   q^2 (s_1+s_2)+(s_1-s_2)^2},\nonumber\\
A_2 =&\frac{1}{(q^4 + (s_1 - s_2)^2 - 2 q^2 (s_1 + s_2))^2}\big[m^4 (q^4 - 2 q^2 (s_1 - 2 s_2) + (s_1 - s_2)^2) \nonumber\\
&+ (6 m_1^4 + q^4 + 
    q^2 (4 s_1 - 2 s_2) + (s_1 - s_2)^2 - 6 m_1^2 (q^2 + s_1 - s_2)) s_2^2\nonumber\\
& + m_2^4 (q^4 + s_1^2 + 4 s_1 s_2 + s_2^2 - 2 q^2 (s_1 + s_2)) \nonumber\\
&- 
 2 m_2^2 s_2 (q^4 - 2 s_1^2 + q^2 (s_1 - 2 s_2) + s_1 s_2 + s_2^2 + 
    3 m_1^2 (-q^2 + s_1 + s_2))\nonumber\\
& - 
 2 m^2 (m_2^2 (q^4 + s_1^2 + s_1 s_2 - 2 s_2^2 + q^2 (-2 s_1 + s_2))\nonumber\\
& + 
    s_2 (-2 q^4 + (s_1 - s_2)^2 + 3 m_1^2 (q^2 - s_1 + s_2) + q^2 (s_1 + s_2)))\big],\nonumber\\
B_2 =&\frac{1}{(q^4 + (s_1 - s_2)^2 - 2 q^2 (s_1 + s_2))^2}\big[m^4 (q^4 + q^2 (4 s_1 - 2 s_2) + (s_1 - s_2)^2)\nonumber\\
& + 
 s_1^2 (6 m_2^4 + q^4 - 2 q^2 (s_1 - 2 s_2) + (s_1 - s_2)^2 - 
    6 m_2^2 (q^2 - s_1 + s_2)) \nonumber\\
&+ 
 m_1^4 (q^4 + s_1^2 + 4 s_1 s_2 + s_2^2 - 2 q^2 (s_1 + s_2))\nonumber\\
& - 
 2 m_1^2 s_1 (q^4 + s_1^2 + s_1 s_2 - 2 s_2^2 + q^2 (-2 s_1 + s_2) + 
    3 m_2^2 (-q^2 + s_1 + s_2))\nonumber\\
& - 
 2 m^2 (m_1^2 (q^4 - 2 s_1^2 + q^2 (s_1 - 2 s_2) + s_1 s_2 + s_2^2) \nonumber\\
&+ 
    s_1 (-2 q^4 + (s_1 - s_2)^2 + 3 m_2^2 (q^2 + s_1 - s_2) + q^2 (s_1 + s_2)))\big],\nonumber\\
C_2 =&\frac{1}{(q^4 + (s_1 - s_2)^2 - 2 q^2 (s_1 + s_2))^2}\big[3 m_1^4 (q^2 - s_1 - s_2) s_2 + m^4 (2 q^4 - (s_1 - s_2)^2 - q^2 (s_1 + s_2))\nonumber\\
& -
  m^2 (-q^6 + q^4 s_1 + q^2 s_1^2 - s_1^3 + q^4 s_2 - 6 q^2 s_1 s_2 + 
    s_1^2 s_2 + q^2 s_2^2 + s_1 s_2^2 - s_2^3\nonumber\\
& + 
    2 m_2^2 (q^4 + q^2 s_1 - 2 s_1^2 - 2 q^2 s_2 + s_1 s_2 + s_2^2) + 
    2 m_1^2 (q^4 + s_1^2 + s_1 s_2 - 2 s_2^2 + q^2 (-2 s_1 + s_2))) \nonumber\\
&+ 
 2 m_1^2 (-s_2 (q^4 - 2 s_1^2 + q^2 (s_1 - 2 s_2) + s_1 s_2 + s_2^2) + 
    m_2^2 (q^4 + s_1^2 + 4 s_1 s_2 + s_2^2 - 2 q^2 (s_1 + s_2)))\nonumber\\
& - 
 s_1 (3 m_2^4 (-q^2 + s_1 + s_2) + 
    2 m_2^2 (q^4 + s_1^2 + s_1 s_2 - 2 s_2^2 + q^2 (-2 s_1 + s_2)) \nonumber\\
&+ 
    s_2 (-2 q^4 + (s_1 - s_2)^2 + q^2 (s_1 + s_2)))\big],\nonumber\\
D_2=& \frac{1}{2 (q^4 + (s_1 - s_2)^2 - 2 q^2 (s_1 + s_2))}\big[m^4 q^2 + m_1^4 s_2 + m_1^2 (m_2^2 (q^2 - s_1 - s_2) \nonumber\\
&+ s_2 (-q^2 - s_1 + s_2)) + 
 s_1 (m_2^4 + q^2 s_2 - m_2^2 (q^2 - s_1 + s_2)) \nonumber\\
&- 
 m^2 (m_2^2 (q^2 + s_1 - s_2) + m_1^2 (q^2 - s_1 + s_2) + q^2 (-q^2 + s_1 + s_2))\big].
\end{align}
\end{widetext}

\section{Analytical Results}\label{app:analyticalRes}

Here we present the analytical results for the calculation of perturbative diagram and $GG$ condensate diagrams. The imaginary part of the correlation function contributed by perturbative diagram reads as
\begin{widetext}
\begin{align}
&{\rm Im}^2 {\tilde \Pi}^{\rm pert}(p_1,p_2,q^2)\nonumber\\
=&\frac{N_c}{4 \pi } I_{\Delta}[m_s^2 (2 A_1+10
   B_1-1)+B_1 (-3
   q^2-s_1+s_2)+s_1].
   \end{align}
The imaginary parts of the correlation function contributed by the $GG$ condensate  diagrams read as
   \begin{align}
&{\rm Im}^2 {\tilde \Pi}^{GG(a)}(p_1,p_2,q^2)\nonumber\\
=&\frac{\langle GG \rangle}{48 \pi} \frac{\partial^2}{\partial M_1^2 \partial M_2^2} \nonumber\\
   &\times I_{\Delta}[M_1^2 (2 A_1+8
   B_1-1)+B_1 (6
   M_2^2+24 m_s^2-7
   q^2-s_1+s_2)+s_1]|_{M_1^2=M_2^2=m_s^2}.\\
&{\rm Im}^2 {\tilde \Pi}^{GG(b)}(p_1,p_2,q^2)\nonumber\\
=&\frac{\langle GG \rangle}{4\pi}\frac{\partial}{\partial M_1^2}I_{\Delta}(A_1+B_1-1)|_{M_1^2=m_s^2}\nonumber\\
 &-\frac{\langle GG \rangle}{96
   \left((-q^2+s_1+s_2)^
   2-4 s_1 s_2\right)^{3/2}} \frac{\partial}{\partial M_1^2} \nonumber\\
   &\times [M_1^2 (q^2-5 s_1+5
   s_2)+m_s^2
   (q^2+s_1-s_2)+q^2 (-3 q^2+7 s_1+3
   s_2)]\nonumber\\
   &\times \theta[(M_1^2)^2 (-s_2)+M_1^2
   (m_s^2
   (-q^2+s_1+s_2)+s_2
   (q^2+s_1-s_2))-
   s_1 \left(m_s^4-m_s^2
   (q^2-s_1+s_2)+q^2 s_2\right)]\nonumber\\
   &\times\theta(s_1-M_1^2)\theta(s_2-m_s^2)|_{M_1^2=m_s^2}\nonumber\\
   &-\frac{\langle GG \rangle}{48
   \left((-q^2+s_1+s_2)^
   2-4 s_1 s_2\right)^{3/2}}\frac{\partial}{\partial M_1^2}\nonumber\\
   &\times[(M_1^2)^2 s_2+M_1^2
   (q^2-s_1-s_2)
   \left(m_s^2-q^2+s_1-s_2\right)+m_s^4 s_1-2
   m_s^2 \left(q^4-q^2
   (s_1+2 s_2)+s_2
   (s_2-s_1)\right)\nonumber\\
   &+q^2
   \left(q^4-2 q^2
   (s_1+s_2)+s_1^2+\text
   {s1} s_2+s_2^2\right)]\nonumber\\
   &\times\frac{s_2^{(2)}-s_2^{(1)}}{|s_2^{(2)}-s_2^{(1)}|}[\delta(s_2-s_2^{(1)})-\delta(s_2-s_2^{(2)})]|_{M_1^2=m_s^2}\nonumber\\
   &-\frac{\langle GG \rangle}{48
   \left((-q^2+s_1+s_2)^
   2-4 s_1 s_2\right)^{3/2}}\frac{\partial}{\partial M_1^2}\nonumber\\
   &\times[s_1 (M_1^2
   (q^2-s_1+s_2)+m_s^2
   (q^2+s_1-s_2)+q^2
   (-q^2+s_1+s_2))]\nonumber\\
   &\times\frac{s_1^{(2)}-s_1^{(1)}}{|s_1^{(2)}-s_1^{(1)}|}[\delta(s_1-s_1^{(1)})-\delta(s_1-s_1^{(2)})]|_{M_1^2=m_s^2}.\\
&{\rm Im}^2 {\tilde \Pi}^{GG(c)}(p_1,p_2,q^2)\nonumber\\
=&\frac{\langle GG \rangle}{8\pi}\frac{\partial}{\partial M_2^2}I_{\Delta}(A_1+B_1-1)|_{M_2^2=m_s^2}\nonumber\\
 &-\frac{\langle GG \rangle}{48
   \left((-q^2+s_1+s_2)^
   2-4 s_1 s_2\right)^{3/2}} \frac{\partial}{\partial M_2^2} \nonumber\\
   &\times[3 M_2^2
   (q^2+s_1-s_2)-
   m_s^2 (q^2+7 s_1-7
   s_2)+q^2 (-3
   q^2+7 s_1+3 s_2)]\theta[-(M_2^2)^2 s_1+m_s^2
   (M_2^2
   (-q^2+s_1+s_2)\nonumber\\
   &+s_2
   (q^2+s_1-s_2))+M_2^2 s_1
   (q^2-s_1+s_2)+m_s^4 (-s_2)-q^2 s_1
   s_2]\theta(s_1-m_s^2)\theta(s_2-M_2^2)|_{M_2^2=m_s^2}\nonumber\\
   &-\frac{\langle GG \rangle}{24
   \left((-q^2+s_1+s_2)^
   2-4 s_1 s_2\right)^{3/2}}\frac{\partial}{\partial M_1^2}\nonumber\\
   &\times[(M_2^2)^2 s_1+M_2^2
   (q^2-s_1-s_2)
   \left(m_s^2-q^2-s_1+
   s_2\right)+m_s^4 s_2-2
   m_s^2 \left(q^4-q^2
   (2 s_1+s_2)+s_1
   (s_1-s_2)\right)\nonumber\\
   &+q^2
   \left(q^4-2 q^2
   (s_1+s_2)+s_1^2+s_1 s_2+s_2^2\right)]\nonumber\\
   &\times\frac{s_2^{(2)}-s_2^{(1)}}{|s_2^{(2)}-s_2^{(1)}|}[\delta(s_2-s_2^{(1)})-\delta(s_2-s_2^{(2)})]|_{M_2^2=m_s^2}\nonumber\\
   &-\frac{\langle GG \rangle}{24
   \left((-q^2+s_1+s_2)^
   2-4 s_1 s_2\right)^{3/2}}\frac{\partial}{\partial M_1^2}\nonumber\\
   &\times[M_2^2
   (q^2+s_1-s_2)+m_s^2
   (q^2-s_1+s_2)+q^2 (-q^2+s_1+s_2)]\nonumber\\
   &\times\frac{s_1^{(2)}-s_1^{(1)}}{|s_1^{(2)}-s_1^{(1)}|}[\delta(s_1-s_1^{(1)})-\delta(s_1-s_1^{(2)})]|_{M_2^2=m_s^2},
\end{align}
where 
\begin{align}
s_1^{(1)}&=\frac{m_1^2
   \left(m_2^2+s_2\right)+\left(m_2^2-s_2\right)
   \left(\sqrt{m_1^4-2 m_1^2
   \left(m_2^2+q^2\right)+\left(m_2^2-q^2\right)^2}-m_2^2+q^2\right)}{2 m_2^2},\nonumber\\
s_1^{(2)}&=\frac{m_1^2
   \left(m_2^2+s_2\right)-\left(m_2^2-s_2\right)
   \left(\sqrt{m_1^4-2 m_1^2
   \left(m_2^2+q^2\right)+\left(m_2^2-q^2\right)^2}+m_2^2-q^2\right)}{2 m_2^2},\nonumber\\
s_2^{(1)}&=-\frac{m_1^4-m_1^2 m_2^2-m_1^2 q^2-m_1^2
   s_1+\left(s_1-m_1^2\right) \sqrt{m_1^4-2 m_1^2
   \left(m_2^2+q^2\right)+\left(m_2^2-q^2\right)^2}-m_2^2
   s_1+q^2 s_1}{2 m_1^2},\nonumber\\
s_2^{(2)}&=-\frac{m_1^4-m_1^2 m_2^2-m_1^2 q^2-m_1^2
   s_1+\left(m_1^2-s_1\right) \sqrt{m_1^4-2 m_1^2
   \left(m_2^2+q^2\right)+\left(m_2^2-q^2\right)^2}-m_2^2
   s_1+q^2 s_1}{2 m_1^2}.
\end{align}
\end{widetext}
It should be noted that for the $GG(a)$ diagram, before doing derivatives on $M_1^2$ and $M_2^2$, one must set the mass of the $k_1, k_2, k$ lines as $M_1, M_2, 0$. For the $GG(b)$ and $GG(c)$ diagrams, before doing derivatives, one must set the mass of the $k_1, k_2, k$ lines as $M_1, m_s, 0$ and $m_s, M_2, 0$, respectively.
\end{appendix}

\end{document}